\documentclass[prl,showpacs,floatfix,twocolumn]{revtex4}
\usepackage{graphicx,graphics,epsfig,psfrag}
\usepackage{dcolumn,bm}
\usepackage{latexsym,amssymb}
\begin{document}

\title{Effects of Scale-Free Disorder on the Anderson Metal-Insulator Transition}

\author{Macleans L. Ndawana}
\altaffiliation[Permanent address: ]{Department of Physics,
University of Zambia, P. O. Box 32379, Lusaka, Zambia}

\author{Rudolf A. R\"{o}mer}
\affiliation{Department of Physics and Centre for Scientific
Computing, University of Warwick, Coventry, CV4 7AL, United
Kingdom}

\author{Michael Schreiber}
\affiliation{Institut f\"{u}r Physik, Technische Universit\"{a}t
Chemnitz, 09107 Chemnitz, Germany}
\date{\today}

\begin{abstract}
  We investigate the three-dimensional Anderson model of localization
  via a modified transfer-matrix method in the presence of scale-free
  diagonal disorder characterized by a disorder correlation function
  $g(r)$ decaying asymptotically as $r^{-\alpha}$. We study the
  dependence of the localization-length exponent $\nu$ on the
  correlation-strength exponent $\alpha$.
  For fixed disorder $W$, there is a critical $\alpha_{\rm c}$, such
  that for $\alpha < \alpha_{\rm c}$, $\nu=2/\alpha$ and for $\alpha >
  \alpha_{\rm c}$, $\nu$ remains that of the uncorrelated system in
  accordance with the extended Harris criterion.  At the band center,
  $\nu$ is independent of $\alpha$ but equal to that of the uncorrelated
  system. The physical mechanisms leading to this different behavior are discussed.
\end{abstract}
\maketitle

%%%%%%%%%%%%%%%%%%%%%%%%%%%%%%% INTRODUCTION %%%%%%%%%%%%%%%%%%%%%%%%%%%%%%%%%%%%%
The successful analysis of the metal-insulator transition (MIT) in the
Anderson model of localization \cite{AbrALR79} has hitherto been limited
to short-range or uncorrelated diagonal disorder. In this Letter we
report the effects of long-range power-law correlated disorder --- so
called scale-free disorder --- on the MIT.  Scale-free disorder is
omnipresent in nature. It is found in many diverse situations in
biological \cite{PenBGH92,IvaAGH01} and in physical
\cite{Isc92,VidMNM96} systems, in city growth patterns \cite{MakABH98}
and in economics \cite{ManS00}. The effects of scale-free disorder on
the critical properties of physical systems have recently received much
renewed attention
\cite{BalP99,Ces94,PruPF99,PruF99,PruPF00,BlaFH02,PraHSS92,BouG90,CaiRSR01,SanMK03}.

For the Anderson model of localization previous investigations of
scale-free disorder have concentrated on the one- (1D)
\cite{MouL98,RusKBH99,RusKBH01} and two-dimensional (2D)
\cite{LiuCX99,LiuLL03} cases. For the 1D Anderson model, it has been
shown that for energies close to the band edge the presence of
scale-free diagonal disorder causes states to be strongly localized
\cite{RusHW98}, while at the band center the states tend to have
localization lengths $\xi$ comparable to the system size
\cite{RusKBH99}. In the 2D Anderson model with scale-free disorder an
MIT of the Kosterlitz-Thouless transition type \cite{LiuCX99,LiuLL03}
has been observed.

Of particular interest to us is the influence of scale-free disorder in
the neighborhood of an MIT when the localization length $\xi$ becomes
sufficiently large. Scale-free disorder can affect the character of the
divergence as shown in Refs.~\cite{WeiH83,Wei84} for the {\em classical}
percolation problem. The principle finding of Refs.~\cite{WeiH83,Wei84}
is that the critical exponent $\nu$, governing the divergence of $\xi$,
in a scale-free disordered classical system can change when the
correlator $g({\bf r}-{\bf r'})$ falls off with distance as a power law,
i.e., $\propto |{\bf r}-{\bf r'}|^{-\alpha}$. The critical exponent of
the classical percolation $\nu=4/3$ crosses over to $\nu=2/\alpha$ for
$\alpha < 3/2$, i.e., when the decay of the correlator is slow enough.

The Harris criterion \cite{Har74,Har83} summaries the effects of
short-range correlated disorder on a critical point. The criterion
states that $\nu$ for the disordered system and the clean system are
identical provided that $d\nu-2 > 0$, with $d$ being the dimensionality
of the system. The inequality is derived by demanding that the
fluctuations of the random potential within a volume given by $\xi$ do
not grow faster than their mean value as the transition is approached.
For power-law correlated potentials, like the above scenario, there is
an extension of this criterion originally suggested in
Ref.~\cite{WeiH83} and further studied in Refs.~\cite{Wei84,CaiRSR01}.
The behavior of $\nu$ in the presence of scale-free disorder is well
described by the {\em extended ~Harris criterion} which can be stated
formally as
\begin{equation}
\label{hari}
\nu=\left \{ \begin{array}{ll}
2/\alpha & {\rm if} \quad \alpha < \alpha_{\rm c} \\
\nu_{0}  & {\rm if} \quad \alpha > \alpha_{\rm c}
\end{array} \right. \quad ,
\end{equation}
where $\nu_0$ is the critical exponent without correlations in the
disorder. Eq.~(\ref{hari}) implies that there is a well-defined critical
value $\alpha_{\rm c}=2/\nu_0$, below which correlations are relevant
and above which correlations are irrelevant.  Numerical studies for 2D
classical percolation \cite{PraHSS92} are indeed in good agreement with
Eq.\ (\ref{hari}). Quite recently, numerical investigations of
long-range correlations in models of 2D quantum-Hall systems
corroborated its validity for the quantum case \cite{CaiRSR01,SanMK03}.
Therefore it appears to be possible that the criterion based on
potential fluctuations can be applied also to quantum phase transitions
such as the Anderson-type MIT, where $\nu$ is determined from the
divergence of the {\em quantum}-localization length at the critical
point. It is the purpose of this Letter to investigate this possibility.

\begin{figure}
\resizebox{0.45\textwidth}{!}{
\includegraphics{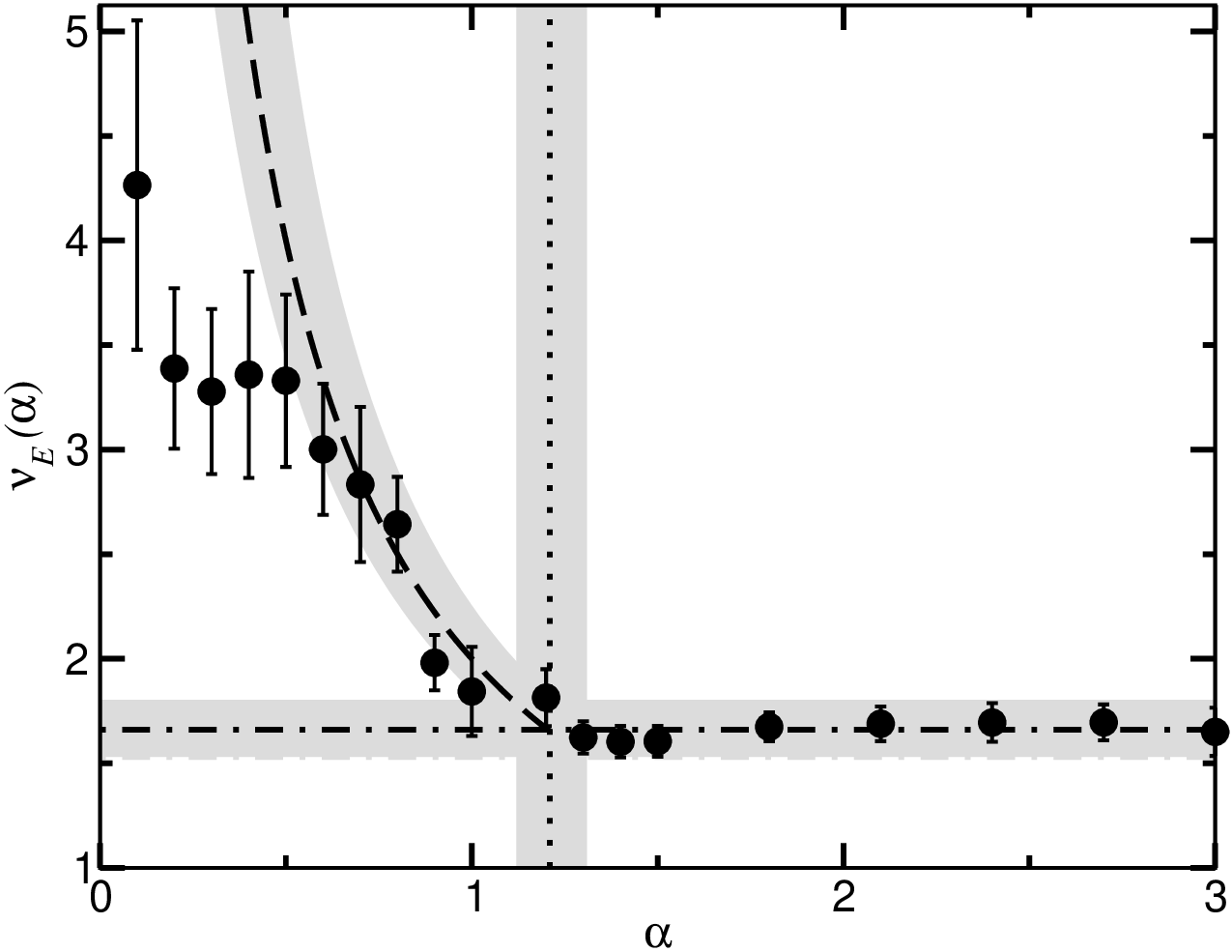}}
\caption{\label{fig-nu_E}
  The localization-length exponent $\nu_{E}(\alpha)$ as a function of
  the correlation-strength exponent $\alpha$ at $W=12$. Error bars
  reflect one standard deviation. The horizontal line indicates the
  uncorrelated $\nu_0=1.66\pm 0.06$, the vertical dotted line is
  $\alpha_{\rm c}=1.21$. The dashed line for $\alpha < \alpha_{\rm c}$
  gives the extended Harris criterion (\ref{hari}). The grey areas
  denote error bounds of one confidence interval arising from the error
  in $\nu_{0}$. Deviations from the extended Harris criterion for small
  $\alpha < 0.5$ are due to finite-size effects.}
\end{figure}

\begin{figure}
\resizebox{0.45\textwidth}{!}{
\includegraphics{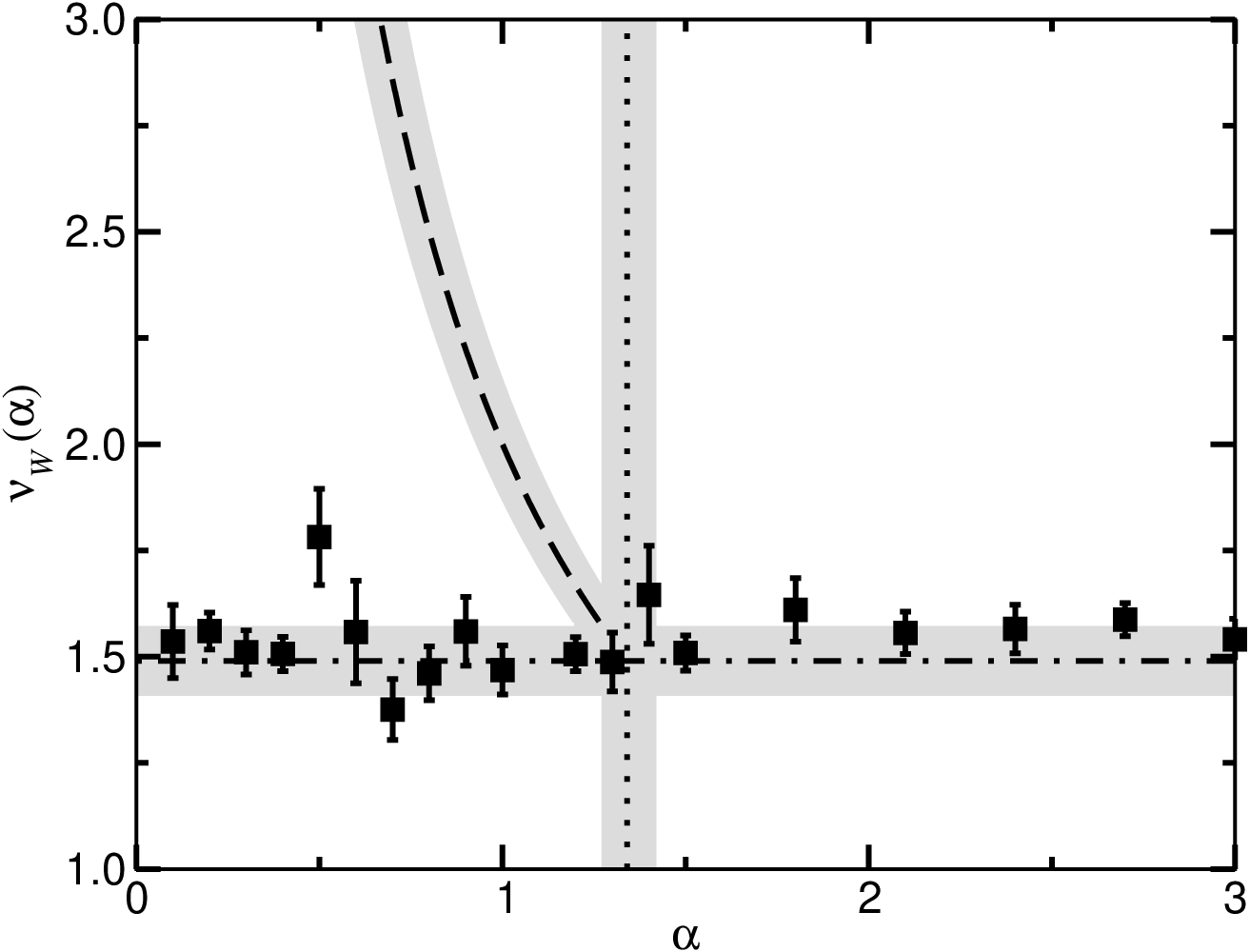}}
\caption{\label{fig-nu_W}
  The localization-length exponent $\nu_{W}(\alpha)$ as a function of
  the correlation-strength exponent $\alpha$ at $E=0$. Error bars
  reflect one standard deviation. The horizontal line indicates the
  uncorrelated $\nu_0=1.49\pm 0.03$, the vertical dotted line is
  $\alpha_{\rm c}=1.34$. The dashed line for $\alpha < \alpha_{\rm c}$
  gives the extended Harris criterion (\ref{hari}). The grey areas
  denote error bounds of one confidence interval arising from the error
  in $\nu_{0}$.}
\end{figure}

We define $\nu_E$ and $\nu_W$ as critical exponents of the localization
length $\xi\propto |E-E_{\rm c}|^{-\nu_E}$ at fixed disorder strength
$W$ and $\xi\propto |W-W_{\rm c}|^{-\nu_W}$ at fixed energy $E$,
respectively. We use the symbol $\nu$ to denote both exponents $\nu_E$
and $\nu_W$. For fixed $W$ and $\alpha < \alpha_{\rm c}$, the critical
exponent $\nu_E$ obeys the extended Harris criterion, as shown in
Fig.~\ref{fig-nu_E}. At the band center $E=0$, however, $\nu_W$ remains
independent of $\alpha$ as can be seen in Fig.~\ref{fig-nu_W}. This
means that scale-free disorder increases the critical exponent $\nu_E$
for small $\alpha$ while leaving $\nu_W$ unchanged.

Our calculation is based upon the Anderson tight-binding Hamiltonian
\cite{And58} in site representation
\begin{equation}
\label{hamit} {\cal H}=\sum_{\langle
i,j\rangle}\left|i\rangle\langle j\right| +
\sum_{i}\varepsilon_{i}\left|i\rangle\langle i\right|
\end{equation}
where $\langle i,j \rangle$ denotes a sum over nearest-neighbors and
$\left| i \rangle\right.$ is an atomic-like orbital at site $i$. The
random on-site potentials $\varepsilon_{i}$ are chosen from a Gaussian
distribution with zero mean and variance $W^{2}/12$. For uncorrelated
Gaussian disorder, the dependence of $W_{c}(E)$ is known \cite{BulSK87},
in particular $W_{c}(0)=20.9 \pm 0.5$. We generate scale-free disorder
by use of the modified Fourier-filtering method (FFM) as outlined in
Refs.~\cite{MakHSS95,MakHSS96}, so that the random on-site energies have
an asymptotic correlation function
$\langle \varepsilon_{i}\varepsilon_{i+r}\rangle \sim r^{-\alpha}$
in real space. The average is done over spatial positions and many
disorder realizations. We note that large $\alpha$ corresponds to the
nearly uncorrelated case which shall serve as our point of reference,
while small $\alpha$ is the strongly correlated case.

In principle the usual iterative transfer-matrix method (TMM)
\cite{PicS81a,PicS81b,MacK83,KraM93,Mac94} allows us to determine the
localization length $\lambda$ of electronic states in a quasi-1D system
with cross section $M \times M$ and length $L \gg M$, where typically a
few million sites are needed for $L$ to achieve a reasonable accuracy
for $\lambda$. However, the use of the non-iterative FFM procedure to
generate scale-free disorder necessitates a complete storage of the
on-site potentials and consequently, the iterative advantage of TMM is
lost as computer memory requirements become rather large.

In order to circumvent this problem, we have modified the conventional
TMM. We now perform the TMM on a system of fixed length $L_0$ of the
quasi-1D bar. After the usual forward calculation with a global transfer
matrix ${\cal T}_{L_0}$, we add a backward calculation with transfer
matrix ${\cal T}^{\rm b}_{L_0}$. This forward-backward-multiplication
procedure is repeated $K$ times. The effective total number of TMM
multiplications is $L_{\rm }=2KL_0$ and the global transfer-matrix
${\tau}_{L_{\rm }}$ is
\begin{eqnarray}
\label{prodtf} {\tau}_{L_{\rm }} & = & \left(T^{\rm b}_{1} \cdots
T^{\rm b}_{n} \cdots T^{\rm b}_{L_0}T_{L_0}\cdots
T_{n}\cdots T_{1}\right)^{K} \nonumber \\
& = & \left( {\cal T}^{\rm b}_{L_0} {\cal T}_{L_0}\right)^K \quad .
\end{eqnarray}
As usual, we diagonalize the matrix
\begin{equation}
\label{osel2} \Gamma_{L_{\rm }} \approx \lim_{K\rightarrow
\infty}\left({\tau}^{\dagger}_{L_{\rm }} {{\tau}_{L_{\rm
}}}\right)^{{1}/{4KL_{0}}} \quad .
\end{equation}
This modified TMM has previously been used for two interacting particles
\cite{RomS97b}.

After establishing Eq.~(\ref{osel2}) the calculation of $\lambda$
follows that for the conventional TMM \cite{remark}. The matrix
$\Gamma_{L_{\rm }}$ is symplectic with $M^2$ paired eigenvalues $\exp
(\pm\gamma)$ with Lyapunov exponents $\gamma$ . Physically, $\gamma$
determines the increase or decrease of the envelope of the wave function
at long distances. The localization length is defined as the inverse of
the smallest Lyapunov exponent, $\lambda=1/\gamma_{\rm min}$
\cite{Bor63,Ose68,CriPV93}. For a reliable convergence check, only the
accumulated changes to $\lambda$ after each complete
forward-and-backward loop need to be taken into account, not all such
changes in the bulk ($n= 2, 3, \ldots, L_0-1$) of the sample. Usually,
convergence can be achieved after just a few $K$ multiplications.

The critical behavior can be determined by numerically establishing for
the reduced-localization lengths $\Lambda=\lambda/M$ the one-parameter
scaling hypothesis \cite{AbrALR79,MacK83,SleO99a,MilRSU00} as
$\Lambda(x,\alpha)={\cal F}[\xi(x,\alpha)/M]$ in the vicinity of the
MIT, where $\xi(x,\alpha)$ is the three-dimensional (3D) localization length in 
the thermodynamic limit. And accordingly \cite{AbrALR79,MacK83}, $\xi$ 
diverges when the tuning parameter $x$, which can be the disorder strength 
$W$ or the energy $E$, is tuned to its critical value $x_{\rm c}$, i.e., 
$W_{\rm c}$ or $E_{\rm c}$ as
\begin{equation}
\label{corrl}
\xi(x) \sim \left|x-x_{\rm c}\right|^{-\nu}.
\end{equation}
The value of $\nu$ is estimated from the one-parameter scaling
hypothesis \cite{AbrALR79,MacK83} by finite-size scaling (FSS)
\cite{SleO99a,SleO99b,MilRSU00}. The FSS procedure performed here
follows closely the approach in
Refs.~\cite{NdaRS02,MilRS99a,MilRS01,KawKO98}. The basic idea
\cite{KraM93} is to construct a family of fit functions which include
corrections to scaling \cite{Mac94} due to an irrelevant scaling
variable and due to non-linearities of the disorder dependence of the
scaling variables \cite{SleO99a}.

%%%%%%%%%%%%%%%%%%%%%%%%%%%%%%%%%% PHASE DIAGRAM %%%%%%%%%%%%%%%%%%%%%%%%%%%%%%%%%
Numerical results have been obtained for samples of lengths up to at
least $L_0=1000$, system widths $M=5,7,9,11,13$ and for various $\alpha$ 
values.  For band center data as in Fig.~\ref{fig-nu_W}, widths up to $M=15$ have 
been used. For each sample, convergence was assumed when the value of 
$\gamma$ changed by less than $10^{-5}$ after a complete forward-and-backward 
loop. We emphasize that since no self-averaging is used, the more stringent
standard TMM-convergence criterion \cite{MacK83} is unnecessary. With at
least $100$ samples for each $\lambda(W,E,\alpha)$, we find a relative
error of the sample-averaged localization lengths of $\lesssim 5\%$.

\begin{figure}
\resizebox{0.45\textwidth}{!}{
\includegraphics{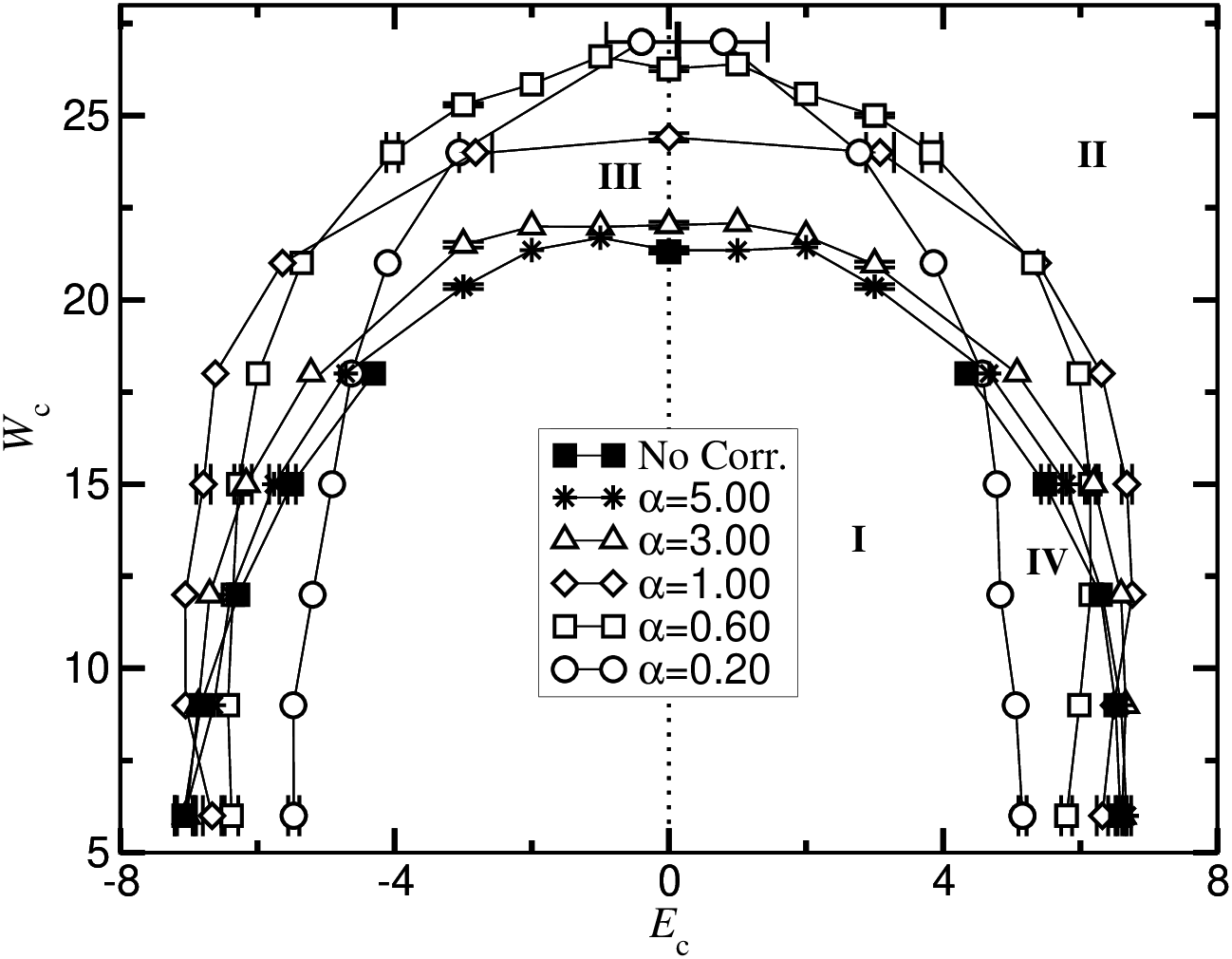}}
\caption{\label{fig-phasediag}
  The phase diagram of the 3D Anderson model of localization for values
  of $\alpha$ shown in the legend. For $\alpha \gg \alpha_{\rm c}$ the
  uncorrelated phase diagram is recovered. For clarity, error bars of one standard
  deviation are shown for every third data point only.}
\end{figure}

The phase diagram of localization, Fig.~\ref{fig-phasediag}, shows the
mobility-edge trajectories, which separate extended from localized
states, for different values of $\alpha$. The symmetry with respect to
$E=0$ holds as in the uncorrelated case. $W_{\rm c}(\alpha)$ at $E=0$
increases monotonically as $\alpha\rightarrow 0$.
In the absence of correlations, states in region I are extended whereas
states in II are localized. The phase boundaries between the metallic
states (region I) and the insulating states (region II) are modified in
the presence of scale-free disorder.
With uncorrelated disorder, states in region III are localized by
quantum interference effects \cite{BulKM85,BulSK87}. Correlation in the
disorder potential will lead to a smoother disorder and a decrease of
interference and thus increased localization lengths.  Indeed, we find
that for $\alpha<\infty$, states in region III become {\em extended} and
the phase boundary is shifted to higher values of the disorder. The
effect becomes most pronounced for strongly correlated disorder, i.e.,
small $\alpha$. For an initially insulating system with
weakly-correlated disorder slightly larger than $W_{\rm c}$, we find a
transition to metallic behavior upon decreasing $\alpha$.
In region IV, the phase boundary moves into the metallic phase I in
contrast to region III. Thus states which are extended for weakly correlated
disorder, become localized for strong correlation $\alpha \rightarrow 0$. As 
has been argued before \cite{BulKM85,BulSK87}, the critical
behavior near the mobility edges is governed by quantum interference as
well as tunneling between potential wells. For small $W$ and large $|E|$
i.e., at the band edge the potential wells govern the localization
behavior \cite{EcoSZ84}. Obviously, a disorder potential smoothened by
correlations will also affect these tunneling processes, and away from
the band center, this results in more localization. A similar effect has
been seen for the 1D case \cite{RusHW98}.

%%%%%%%%%%%%%%%%%%%%%%% CRITICAL EXPONENTS %%%%%%%%%%%%%%%%%%%%%%%%%%%%%%%%%%%%%%%

The different influence of scale-free disorder on the states at the band
center and at the band edge is also the origin of the difference in the
behavior of $\nu_{E}(\alpha)$ and $\nu_{W}(\alpha)$ shown in Figs.\
\ref{fig-nu_E} and \ref{fig-nu_W}. The extended Harris criterion
essentially states that the critical behavior of the MIT remains
unaffected if, as $E\rightarrow E_{\rm c}$, the divergence of the
localization length $\xi \sim |E-E_{\rm c}|^{-\nu_{E}}$ is stronger than
the divergence of a typical size $|E-E_{\rm c}|^{-2/\alpha}$ associated
with a correlated disorder potential fluctuation of energy $|E-E_{\rm
  c}|$.  Otherwise, $\nu_E = 2/\alpha$ \cite{WeiH83,Wei84}.
Let us approximate $E_{\rm c}(W)$ close to $W_{\rm c}$ as
\begin{equation}
\label{rob0} E_{\rm c}(W) \approx |W-W_{\rm c}|^{\beta}.
\end{equation}
Then at $E=0$
\begin{equation}
\label{rob2} \xi\sim |E-E_{\rm c}|^{-\nu_E} = |E_{\rm c}|^{-\nu_E}
\approx |W-W_{\rm c}|^{-\beta\nu_E},
\end{equation}
so that the critical exponents should be related as
\begin{equation}
\label{rob3} \nu_W \approx \beta \nu_E \quad .
\end{equation}
{}From Fig.~\ref{fig-phasediag}, we see that $\beta \ll 1$ in the band
center whereas $\beta=1$ is possible for $|E|>0$.  Furthermore, the
above Harris-type argument is based on classical potential fluctuations,
whereas at $E=0$ the physics is dominated by quantum interference with
an unchanged universal Anderson exponent $\nu_W \approx 1.6$
\cite{SleO99a}.

%%%%%%%%%%%%%%%%%%%%%%%%% SUMMARY %%%%%%%%%%%%%%%%%%%%%%%%%%%%%%%%%%%%%%%%%%%%%%%%
In summary, we have shown that the extended Harris criterion is obeyed
varying $E$ at fixed $W$. The resulting exponent $\nu_E$ agrees very
well with the predictions of the extended Harris criterion. This is the
first such demonstration for a fully quantum coherent situation in three
dimensions to the best of our knowledge.
Moreover, we show that the extended Harris criterion fails at the band
center and trace this failure to the different mechanisms governing the
MIT in the vicinity of the band center and outside.  We emphasize that
such different scaling behavior at the band center and at the band edges
has indeed been speculatively discussed for a long time
\cite{BulKM85,BulSK87}, although numerical studies of $\nu_E$ and
$\nu_W$ for the uncorrelated case suggest a common value. Here we show
that suitably long-ranged power-law correlations with $\alpha <
\alpha_{\rm c}$ give rise to a difference in $\nu_E$ and $\nu_W$.

%%%%%%%%%%%%%%%%% ACKNOWLEDGEMENTS %%%%%%%%%%%%%%%%%%%%%%%%%%%%%%%%%%%%%%%%%%%%%%%
It is a pleasure to thank R.C.\ Ball for stimulating discussions.  This
work is supported by the Deutsche Forschungsgemeinschaft via the SFB 393
and in part via the priority research program "Quanten-Hall-Systeme".
%%%%%%%%%%%%%%%%%%%%%%%%%%%%%%%%%%%%%%%%%%%%%%%%%%%%%%%%%%%%%%%%%%%%%%%%%%%%%%%%%%


\begin{thebibliography}{10}
\bibitem{remark}
Due to the finite size of the samples, special attention is needed
to ensure that the correct mean and variance for the correlated
disorder are used, especially at small $\alpha$.
\end{thebibliography}
\end{document}